\documentclass[nofootinbib,twocolumn,eqsecnum,floats,aps]{revtex4}

\usepackage{bm}
\usepackage{longtable}
\newcommand{\ie}{\emph{i.e.}}
\newcommand{\p}{\mathsf{p}}
\newcommand{\jp}{\mathsf{j}_{p}}
\newcommand{\jL}{\mathsf{j}_{\Lambda}}
\newcommand{\roundbra}[1]{\left( #1 \right|}
\newcommand{\roundket}[1]{\left| #1 \right)}
\newcommand{\phf}{\mathrm{p.h.f.}}

\def\bit{\begin{itemize}}
\def\eit{\end{itemize}}
\def\bnu{\begin{enumerate}}
\def\enu{\end{enumerate}}

\def\e {{\epsilon}}

\def\A {{{\cal A}}}
\def\B {{{\cal B}}}

\def\O {{{\cal O}}}
\def\R {{{\cal R}}}

\def\nn{\nonumber }

\def\M {{{\cal M}}}
\def\x{\times}
\def\Ket#1{||#1 \rangle}
\def\Bra#1{\langle #1||}

\def\ie{{\em i.e., }}
\def\cf{{\it cf. }}
\def\nn{\nonumber }
\def\be{\begin{equation}}
\def\ee{\end{equation}}
\def\br{\begin{eqnarray}}
\def\er{\end{eqnarray}}
\def\brn{\begin{eqnarray*}}
\def\ern{\end{eqnarray*}}
\def\etc{ {\it etc}}

\def\e {{\epsilon}}

\def\bra#1{\langle #1|}
\def\ket#1{|#1 \rangle}
\def\rf#1{{(\ref{#1})}}

\def\sixj#1#2#3#4#5#6{\left\{\negthinspace\begin{array}{ccc}
#1&#2&#3\\#4&#5&#6\end{array}\right\}}
\def\ninj#1#2#3#4#5#6#7#8#9{\left\{\negthinspace\begin{array}{ccc}
#1&#2&#3\\#4&#5&#6\\#7&#8&#9\end{array}\right\}}

\def\go{\rightarrow  }

\def\fot{\frac{1}{2}}
\def\k {{\kappa}}
\def\J {{{\cal J}}}

\def\isim{\:\raisebox{-0.5ex}{$\stackrel{\textstyle.}{=}$}\:}
\begin{document}
\title{s-Wave Approximation
for  Asymmetry  in  Nonmesonic Decay of Finite Hypernuclei}

\author{Cesar Barbero$^{1,4}$}
\author{Alfredo P. Gale\~ao$^2$}
\author{Francisco Krmpoti\'c$^{3,4,5}$}

\affiliation{$^1$Facultad de Ciencias Exactas, Departamento de
F\'isica, Universidad Nacional de La Plata, 1900 La Plata,
Argentina}

\affiliation{$^2$ Instituto de F\'{\i}sica Te\'orica,
Universidade Estadual Paulista \\
Rua Pamplona 145, 01405-900 S\~ao Paulo, SP, Brazil}

\affiliation{$^3$Departamento de F\'isica Matem\'atica, Instituto
de F\'isica da Universidade de S\~ao Paulo, Caixa Postal 66318,
05315-970 S\~ao Paulo, SP, Brazil}

\affiliation{$^4$Instituto de F\'isica La Plata, CONICET, 1900 La
Plata, Argentina}

\affiliation{$^5$Facultad de Ciencias Astron\'omicas y
Geof\'isicas, Universidad Nacional de La Plata, 1900 La Plata,
Argentina}

\date{\today}

\begin{abstract}

We establish the bridge between the commonly used
Nabetani-Ogaito-Sato-Kishimoto (NOSK)   formula for the asymmetry
parameter $a_\Lambda$ in the  $\vec\Lambda p \rightarrow np$
emission of polarized  hypernuclei, and the
 shell model (SM) formalism for finite hypernuclei.
We demonstrate that  the s-wave approximation leads to a SM
formula   for $a_\Lambda$ that is as simple as the NOSK one, and
that  reproduces the exact results for $^5_\Lambda$He and
$^{12}_\Lambda$C better than initially expected. The simplicity
achieved here is indeed remarkable. The new formalism makes the
theoretical evaluation of  $a_\Lambda$ more transparent, and
explains clearly why the one-meson exchange model is unable to
account for the  experimental data of $^5_\Lambda$He.

\end{abstract}


\maketitle

\section{Introduction \label{Int}}

 Despite recent important developments in the  $\Lambda N
 \rightarrow N N$ nonmesonic weak decay (NMWD)  \cite{Al02}, its reaction
mechanism is not fully understood. Indeed, an open problem concerns
the asymmetry parameter $a_\Lambda$ in the  $\vec\Lambda p
\rightarrow np$  emission of the polarized hypernuclei $^5_\Lambda\vec He$ and
 $^{12}_{\phantom{1}\Lambda}\vec{C}$, which yields information on the interference
 between parity conserving (PC) and  parity violating  (PV) transitions.
The measurements favour $a_\Lambda(^5_\Lambda\vec He)>0 $ and
$a_\Lambda(^{12}_\Lambda\vec C)<0 $~\cite{Aj92}, while the
calculations yield almost the same negative value for both
\cite{Al02,Ba05}.

The intrinsic  $\Lambda$ asymmetry is  usually evaluated
from   the formula,
\begin{equation}\label{1.1} a_\Lambda=2\sqrt{3}\,\frac{\Re[
 ae^* -b\, ( c - \sqrt{2}\, d )^*/\sqrt{3}\, +  f\, (
\sqrt{2}\, c + d )^*]} { |a|^2 + |b|^2 + 3 \left( |c|^2 + |d|^2 +
|e|^2 + |f|^2 \right) } \,,
\end{equation}
where the two-body  $|\Lambda p;lSJ\rangle
\rightarrow|np;l'S'J\rangle$ nonmesonic  $p\Lambda\go pn$ decay  amplitudes
\begin{equation}\label{1.2}
\begin{array}[b]{lll}
a = \bra{ ^1\!\mathrm{S}_0}{\hat V}\ket{ ^1\!\mathrm{S}_0},
\quad
&
b = \bra{ ^3\!\mathrm{P}_0}{\hat V}\ket{ ^1\!\mathrm{S}_0},
\quad
&
c = \bra{ ^3\!\mathrm{S}_1}{\hat V}\ket{ ^3\!\mathrm{S}_1},
\\
& & \\
d = \bra{ ^3\!\mathrm{D}_1}{\hat V}\ket{ ^3\!\mathrm{S}_1},
\quad
&
e = \bra{ ^1\!\mathrm{P}_1}{\hat V}\ket{ ^3\!\mathrm{S}_1},
\quad
&
f = \bra{ ^3\!\mathrm{P}_1}{\hat V}\ket{ ^3\!\mathrm{S}_1}
\end{array}
\end{equation}
are the
kinematical correspondents of the inverse reaction $pn\go p\Lambda$.
Eq. \rf{1.1}  was  derived  by  Nabetani, Ogaito,  Sato
and  Kishimoto (NOSK) \cite{Na99} considering only the s-wave
production for the $ p\Lambda$ final states. This
s-wave approximation (s-WA) can be used for the NMWD
straightforwardly  only in the context of the Fermi gas
 model  (FGM),
where  the $\Lambda$-hyperon is embedded in the infinite nuclear matter,
and  is taken to be always in a relative
$s$-state with respect to any of the nucleons within the
hypernucleus~\cite{Du96}.
\footnote{ Note that
the FGM expression
~\cite[Eq. (88)]{Du96} for
 $a_\Lambda$
is incomplete since  it covers
 only the last term in the numerator of Eq. \rf{1.1}.}
 There are several differences
between the scattering states $p\Lambda$ in the reaction $pn\go
p\Lambda$, and  the shell model (SM) description of the nuclear bound
states $p\Lambda$. In the SM the hyperon
stays in the $1s_{1/2}$ orbital, and depending on the
hypernucleus, the proton can occupy the orbitals
 $1s_{1/2}$, $1p_{3/2}$, $1p_{1/2},\cdots$. It is true that  in the
 case of $1s_{1/2}$-shell hypernuclei, the initial $p\Lambda$ system can
 be assumed to be in the relative s-wave state and therefore it
is sufficient to consider only the six matrix elements \rf{1.2}.
 In fact, following the Block-Dalitz anzatz~\cite{Bl63}
for the employment of the FGM in finite nuclei, one can use  the
Eq. \rf{1.1}
 for $^5_\Lambda$He \cite{Sa05}.
But  in the case of
$^{12}_{\phantom{1}\Lambda}{C}$ both the $1s_{1/2}$ and $1p_{3/2}$
single-particle states contribute and, in addition to  the
relative s-state, one has to consider the relative p-state as well
 \cite{Ba02,Kr03,Ba03,Ba05}.

Based on the above arguments, in our previous work \cite{Ba05}, we
 have derived a SM expression   for $a_\Lambda $
which is valid for  both
$^5_\Lambda$He and $^{12}_{\phantom{1}\Lambda}$C. The NOSK
formula  \rf{1.1} has been used there
only for the sake of numerical comparison in the case of
$^5_\Lambda$He. Here we go a step further, establishing a
bridge between the two formalisms. More specifically, we
show that  under plausible
assumptions, the  s-WA can  also be introduced in the SM, yielding
 a NOSK-like formula, which
can be used in the finite hypernuclei
$^5_\Lambda$He and $^{12}_{\phantom{1}\Lambda}$C.

\section{Exact expression for $a_\Lambda$ \label{Exa}}

To introduce the notation, we give here a short account of the formalism we have developed for the calculation
of $a_\Lambda$ in Ref. \cite{Ba05}, where more details can be found.

The mixed state of a hypernucleus having vector polarization $\bm{P}_V$ can be represented by the density matrix \cite[Eq.(9.29)]{Au70}
\begin{equation} \label{2.1}
\rho(J_I) = \frac{1}{2J_I+1}\left[ 1 + \frac{3}{J_I+1}
\bm{P}_V \cdot  \bm{J}_I  \right] \,,
\end{equation}
where $J_I$ is the hypernuclear spin.
The angular distribution of primary protons emitted by such a
hypernucleus is then given by
\begin{eqnarray}
&&\nonumber\lefteqn{
\frac{d\Gamma[\rho(J_I) \to \hat{\bm{p}}_2t_p]}{d\Omega_{p_2}}
= \int d\Omega_{p_1} \int dF\, \sum_{s_1s_2M_F} \sum_{M_IM'_I}}
\\ \nonumber & \times &
\bra{\bm{p}_1 s_1 t_n\, \bm{p}_2 s_2 t_p\,\nu_F J_FM_F} V \ket{J_IM_I}
\bra{J_IM_I}\rho(J_I)\ket{J_IM'_I}
\\ & \times &
\bra{J_IM'_I} V^\dagger
\ket{\bm{p}_1 s_1 t_n\, \bm{p}_2 s_2 t_p\,\nu_F J_FM_F},
\label{2.2}\end{eqnarray}
where $V$ is the nonmesonic transition potential,
$\bm{p}_1 s_1 t_n\equiv -1/2$ and $\bm{p}_2 s_2 t_p\equiv +1/2$ are the momenta and spin
and isospin projections of the emitted neutron and proton, respectively, and $\ket{\nu_FJ_FM_F}$
are the possible final states of the residual nucleus,
where $\nu_F$ specifies the remaining quantum numbers besides those related to the nuclear spin.
Furthermore, we have introduced the compact notation
\begin{eqnarray}
\lefteqn{
\int dF \dots \;=\; 2\pi\, \sum_{\nu_FJ_F}
\int \frac{p_2^2\, dp_2}{(2\pi)^3} \int \frac{p_1^2\, dp_1}{(2\pi)^3}\;
}
\\ \nonumber & \times &
\delta\left( \frac{p_1^2}{2M_N} + \frac{p_2^2}{2M_N}
+ \frac{|\bm{p}_1 + \bm{p}_2|^2}{2M_R}
- \Delta_{\nu_FJ_F} \right) \dots \;,
\label{2.3}
\end{eqnarray}
where the delta function enforces energy conservation,
$M_R$ is  mass  of the  residual nucleus, and $\Delta_{\nu_FJ_F}$ is
the liberated energy.

It is possible to show that the Eq.~\rf{2.2} can be put in the form
\begin{equation}\label{2.4}
\frac{d\Gamma[ \rho(J_I) \to \hat{\bm{p}}_2 t_p ]}{d\Omega_{p_2}} =
\frac{\Gamma_p}{4\pi}
\, \left( 1 + A_V\,\bm{P}_V \cdot  \hat{\bm{p}}_2  \right) \,,
\end{equation}
where $\Gamma_p$ is the full proton-induced decay rate,
and $A_V$ is the \emph{vector hypernuclear asymmetry}, given by
\begin{equation}\label{2.5}
A_V = \frac{3}{J_I+1}\; \frac{\sum_{M_I} M_I \sigma( J_I M_I)}
{\sum_{M_I}  \sigma( J_I M_I)} \,.
\end{equation}
The new quantities introduced above are the \emph{decay strengths},
\begin{eqnarray}
\lefteqn{
\sigma( J_I M_I) \;=\; \int d\Omega_{p_1} \int dF
\sum_{s_1 s_2 M_F}
}\label{2.6}
\\  &\times&
 \left|
\bra{\bm{p}_1 s_1 t_n\, \bm{p}_2 s_2 t_p\, \nu_FJ_FM_F} V
\ket{J_IM_I}_\phf
\right|^2  \,,
\nonumber
\end{eqnarray}
where the subscript $\phf$ indicates that the transition amplitude must be computed in the proton helicity frame.

To proceed, one must write the transition amplitudes in
Eq.~\rf{2.6} in terms of the two-body matrix elements for the
elementary process $\Lambda p \to np$ occurring between the
appropriate bound $\Lambda p$ states in
the hypernucleus and the allowed free final $np$ states.
To this end it is convenient to work
 in the total spin $(S,M_S)$ and isospin $(T,M_T)$ basis,
and to  change the representation to relative and total momenta,
given respectively by $\bm{p}=(\bm{p}_2 - \bm{p}_1)/2$ and
$\bm{P}= \bm{p}_1 + \bm{p}_2$.
Dropping the $M_T=0$ labels, one obtains
\begin{eqnarray}\label{2.7}
\sigma( J_I M_I) &=&
\int d\Omega_{p_1} \int dF\,
\sum_{SM_SM_F} \label{strength1}\\
&\times& \left|\sum_{T}(-)^T
\bra{\bm{p}  \bm{P} SM_ST\, \nu_F J_FM_F} V \ket{ J_IM_I}_\phf
\right|^2 \,.\nonumber
\end{eqnarray}

Next, we: i) expand the final state in terms
of the relative ($\bm{l}$) and center-of-mass (${\bm L}$)
partial waves of the emitted nucleons  \cite[(2.5)]{Ba02},
ii) drop the subscript $\phf$ due to the rotational invariance
of $V$, and iii) integrate on the angle $\phi_{p_1}$, to obtain
\begin{eqnarray}
\sigma(J_I M_I) &=& \frac{ (4\pi)^5}{2}
\int d\cos\theta_{p_1} \int dF\,
\sum_{SM_SM_F} \,
\nn\\
&\times&\biggl|
\sum_{lL\lambda JT} (-)^T\, i^{-l-L}\,
[Y_{l}(\theta_p,\pi) \otimes Y_{L}(\theta_P,0)]_{\lambda\mu}
\nonumber \\ &\times&
(\lambda \mu SM_S|JM_J) (JM_JJ_FM_F|J_IM_I)
\nonumber \\ &\times&
\bra{plPL\lambda SJT\nu_FJ_F;J_I}V\ket{J_I} \biggr|^2 \,,
\label{2.8}
\end{eqnarray}
where ${\bm \lambda}={\bm l}+{\bm L}$ and ${\bm J}={\bm \lambda}+{\bm S}$, and
\begin{eqnarray}
4p^2 &=& p_1^2 + p_2^2 - 2 p_1 p_2 \cos \theta_{p_1} \,,
\nonumber \\
P^2 &=& p_1^2 + p_2^2 + 2 p_1 p_2 \cos \theta_{p_1} \,,
\nonumber \\
\cos \theta_p &=& \frac{p_2 - p_1\cos\theta_{p_1}}{2p} \,,
\nonumber \\
\cos \theta_P &=&  \frac{p_2 + p_1\cos\theta_{p_1}}{P} \,,
\label{2.9}
\end{eqnarray}
where $\theta_{p_1}$ is the angle that $\bm{p}_1$ makes with  $\bm{p}_2$.

Afterwards, we rewrite  the Eq. \rf{2.5} for $A_V$ in terms of
 the \emph{decay moments}
\begin{eqnarray}
\sigma_0(J_I) &=& \sum_{M_I}  \sigma( J_I M_I) \,,
\nn\\
\sigma_1(J_I) &=& \frac{1}{\sqrt{J_I(J_I+1)}} \sum_{M_I} M_I \sigma( J_I M_I) \,,
\label{2.10}
\end{eqnarray}
 as
\begin{equation}\label{2.11}
A_V = 3\; \sqrt{\frac{J_I}{J_I+1}}\; \frac{\sigma_1(J_I)}{\sigma_0(J_I)}
\end{equation}

We remark that the summations
on $M_S$, $M_F$ and $M_I$ have been explicitly performed in Ref.
\cite{Ba05} (\cf~Eqs. (21) and (27) in that reference).

Moreover, we adopt here both: i) the weak-coupling
model (WCM), where for the  $^{A-1}Z$ core ground state
$\ket{J_C}$, the initial state is: $\ket{J_I}\equiv \ket{(\jL J_C)J_I}$, and ii) the
extreme shell model (ESM), where $\ket{\nu_FJ_F} \equiv
\ket{(\jp^{-1}J_C)J_F}$. $\jL\equiv n_\Lambda\,l_\Lambda\,j_\Lambda$
and $\jp\equiv n_p\,l_p\,j_p$  are the single-particle states for the lambda and proton, respectively.
Under these circumstances, and when the
single-proton sub-shells  are completely filled in $\ket{J_C}$, as
happens in the case of $^{5}$He and  $^{12}$C, one gets:
\begin{eqnarray}
\lefteqn{
\bra{plPL\lambda SJT\nu_FJ_F;J_I}V\ket{J_I}}
\nonumber \\ &=&
(-)^{J_C-J_F+j_p}
\hat{J}\hat{J}_F
\sixj{J_C}{J_I}{j_\Lambda}{J}{j_p}{J_F}.
\nonumber \\ &\times&
\M(plPL\lambda SJT;\Lambda \p)
\label{2.12}
\end{eqnarray}
with
\begin{eqnarray}
&&\M(plPL\lambda SJT;\Lambda \p)
=\fot\left[1-(-)^{l+S+T}\right]
\nonumber \\ &\times&
(-)^{T+1}
({plPL\lambda SJT}|V|{\jL \jp JT}),
\label{2.13}
\end{eqnarray}
where the compact notation $\Lambda\equiv \jL,t_\Lambda=-1/2$, $\p\equiv \jp,t_p$
has been used, and the isospin coupling $\ket{t_\Lambda t_p}=1/\sqrt{2}(\ket{T=1,M_T=0}-\ket{T=0,M_T=0})$
has been carried out.
%

Within the above  description, the liberated energies are independent of  $J_F$, \ie $\Delta_{\nu_FJ_F} \to
 \Delta_{\jp}=M_\Lambda-M_N+\e_{\jL}+\e_{\jp}$,
where the  $\e$'s are separation energies, and $M_\Lambda$ is the hyperon mass.
Forthwith, we rewrite  the integration in Eq.~\rf{2.8} as
\begin{equation} \label{2.14}
\int d\cos\theta_{p_1}\int dF \dots \;=\; \frac{1}{(2\pi)^5} \sum_{\jp} \int dU_{\jp} \sum_{J_F} \dots \;,
\end{equation}
where
\br \label{2.15}
&&\int dU_{\jp}\cdots \;=\; \int d\cos\theta_{p_1}
\int p_2^2\, dp_2 \int p_1^2\, dp_1\;
\nn\\
&\times&
\delta\left( \frac{p_1^2}{2M_N} + \frac{p_2^2}{2M_N}
+ \frac{|{\bf p}_1 +{\bf p}_2|^2}{2M_R}
- \Delta_{\jp} \right)\cdots,
\er
Putting all this together, and performing the summation on $J_F$,
we end up with the decay moments
$\sigma_\kappa(J_I)$, given by \cite[Eq.(34)]{Ba05},
 that have a purely kinematical dependence on
the hypernuclear spin $J_I$. This dependence can be
eliminated within the WCM by defining \cite{Ra92} the
intrinsic asymmetry parameter
\begin{eqnarray}
a_\Lambda&=&
\left\{
\begin{array}{ccc}
A_V&
\hspace{0.5em}\mbox{for}\hspace{0.5em}&J_I=J_C+1/2 \,, \\
-\frac{J_I+1}{J_I}A_V&
\hspace{0.5em}\mbox{for}\hspace{0.5em}&J_I=J_C-1/2 \,,
\end{array}\right.
\label{2.16}
\end{eqnarray}
which in the formalism explained above takes the form \cite{Ba05}:
\begin{equation}\label{2.17}
a_\Lambda = \frac{\omega_1}{\omega_0} \,,
\end{equation}
with  the {decay moments}
\begin{eqnarray}
\omega_\kappa &=&
(-)^\kappa \, \frac{8}{\sqrt{2\pi}} \, \hat{\kappa}^{-1} \,\sum_{\jp}
\int dU_{\jp} \,
Y_{\kappa 0}(\theta_p,0) \nonumber \\ &\times&
\sum_{TT'}(-)^{T+T'} \sum_{LS} \; \sum_{l\lambda J} \;
\sum_{l'\lambda' J'} i^{l-l'}\;
 \nonumber \\ &\times&(-)^{\lambda+\lambda'+S+L+j_p+\frac{1}{2}}
\hat{l}\hat{l}'\hat{\lambda}\hat{\lambda'}\hat{J}^2\hat{J'}^2 \,
(l0l'0|\kappa 0) \, \label{2.18} \\ &\times&
\sixj{\kappa}{1/2}{1/2}{j_p}{J}{J'}
\sixj{\kappa}{J'}{J}{S}{\lambda}{\lambda'}
\sixj{l'}{l}{\kappa}{\lambda}{\lambda'}{L} \nonumber \\ &\times &
\mathcal{M}(plPL\lambda SJT;\Lambda \p)
\mathcal{M}^*(pl'PL\lambda' SJ'T';\Lambda \p), \nonumber
\end{eqnarray}
where $\hat{J}=\sqrt{2J+1}$, \etc. We note that the moments $\omega_\kappa$
do not depend on the hypernuclear spin $J_I$, and that
$L=0$ for the  $1s_{1/2}$ state, and  $L=0$ and $1$ for the
$1p_{3/2}$ state.


 To evaluate the matrix elements in \rf{2.13} one has to carry out the $jj-LS$
recouping and  the Moshinsky transformation \cite{Mo59} on the
ket $|{\jL \jp JT})$ (see \cite[Eq.(2.14)]{Ba02})
 to get
\begin{widetext}
 \br
({plPL\lambda SJT}|V|{\jL \jp JT}) &=&
\hat{j}_\Lambda\hat{j}_p \sum_{\lambda'
S'{\sf nlNL}} \hat{\lambda'}\hat{S'} \ninj{l_\Lambda}
{\fot}{j_\Lambda}{l_p}{\fot}{j_p}{\lambda'}{S'}{J}(PL|{\sf NL})
\nn\\
&\x& ({\sf nlNL}\lambda'|n_\Lambda l_\Lambda n_pl_p\lambda')
\roundbra{p,lL\lambda SJ;T}V \roundket{{\sf nlL} \lambda' S'J;T},
\label{2.19}\er
\end{widetext}
where $({\sf n }\cdots|n_\Lambda\cdots )$ are the Moshinsky brackets
\cite{Mo59}. Here,   ${\sf l}$ and ${\sf L}$ stand for  the
quantum numbers of the  relative and  c.m. orbital angular momenta
in the $\Lambda N$ system. Moreover,
 \be
 (PL|{{\sf NL}}) =\delta_{L,{\sf L}}\int R^2dRj_L(PR)\R_{{\sf NL}}(R),
 \label{2.20}\ee
is the overlap of the center of mass radial
wave functions.
 One is interested here in the $\jp=1s_{1/2}$ state, for which is
$\mathsf{l}=\mathsf{L}=0$, and in the $\jp=1p_{3/2}$ state, for which
 both $\mathsf{l}=0$, $\mathsf{L}=1$, and $\mathsf{l}=1$,
$\mathsf{L}=0$ terms  contribute.

\section{Approximate expression for $a_\Lambda$ \label{App}}

We start this section by neglecting   the kinematical and nonlocal
effects on the NMWD introduced in Ref.~\cite{Ba03}, which, as shown
there and confirmed in
Ref. \cite{Ba05}, do not affect the final results by more than
10-20\%
Afterwards we write the Eq. \rf {2.18} in the form:
\br
\omega_\kappa
&=&\frac{8}{\sqrt{\pi}}\sum_{\jp{\sf lL}}\int dU_{\jp}
Y_{\kappa 0}(\theta_p,0)\O(P;{\sf L}){\cal I}_\kappa(p;\jp{\sf l}).
\nn\\
\label{3.1}\er
where $\O(P;{\sf L})\equiv  (P{\sf L}|{{1{\sf L}}})^2$, and
\be
\O(P;0) = \sqrt{\frac{\pi}{2}}\, b^3 e^{-(Pb)^2/2} \,;
\O(P;1) = \frac{(b P)^2 }{3} \O(P;0),
\label{3.2}
\ee
 with  $b$ being  the oscillator length~\cite{Ba02}.

From Eqs. \rf{2.18}, \rf{2.13} and \rf{2.19}
one sees that the just introduced quantities ${\cal I}_\kappa(p;\jp{\sf l})$
are complicated function of $p$, $\jp$ and ${\sf l}$.
 They  involve  several Racah coefficients and many  summations
on different  angular momenta and isospins. However, after performing all the algebra analytically,
we have demonstrated that {\em the nuclear amplitudes  ${\cal I}_\kappa(p;\jp{\sf l}=0)$ do not depend on $\jp$} , \ie
\br
{\cal I}_\k(p;\jp=1s_{1/2},0)&=&{\cal I}_\k(p;\jp=1p_{3/2},0)\equiv {\cal I}_\k(p;0).
\nn\\\label{3.3}\er
On the other hand for ${\sf l}=1$ only the
 $\jp=1p_{3/2}$ state contributes, and one can write
\br
{\cal I}_\k(p;\jp=1p_{3/2},1)&\equiv& {\cal I}_\k(p;1).
\label{3.4}\er

The explicit expressions for the form quantities ${\cal I}_\kappa(p;{\sf l})$,
in the one-meson-exchange model (OMEM), that comprises the
($\pi,\eta,K,\rho,\omega,K^*$) mesons,
are  exhibited in the Appendix~\ref{A}.
The $\kappa=0$ pieces of  Eq. \rf{3.1}, \ie the Eqs. \rf{A1} and
\rf{A3},
have  already been derived
in a previous work \cite{Ba02}, where we have also learned that the
 matrix elements  contained
within  $ {\cal I}_0(p;1)$
 represent  the higher order contributions (HOC),
when compared with those contained
within  $ {\cal I}_0(p;0)$. These HOC are
 $\cong 2\%$
for the PC transitions and $\cong  15\%$ for the PV transitions
(see also Ref.~\cite{It02}). Here we have verified  numerically
that the HOC  contribute   to $\omega_1$ in similar
proportions, and therefore their overall  effect on
$a_\Lambda$  is
relatively small.
Thus, the ${\sf l}=1$ contributions to $^{12}_\Lambda$C,
will be omitted from now on, and we end up with

\br
\omega_\kappa
&=&\frac{8}{\sqrt{\pi}}\sum_{\jp{\sf L}}\int dU_{\jp}
Y_{\kappa 0}(\theta_p,0)\O(P;{\sf L}){\cal I}_\kappa(p;0),
\label{3.5}\er
which, together with \rf{2.17}  is what we call the s-WA for $a_\Lambda$ in finite
hypernuclei.
Needless to say that  the Eq.  \rf{3.5} is exact for $^{5}_\Lambda$He and
equivalent to \rf{2.18}.  Note that
the summation on  $\jp$ in \rf{3.5}
 only affects the range of the  integration as indicated in \rf{2.15}.

Next we  show that the amplitudes
${\cal I}_0(p;{\sf l}=0)$ and
${\cal I}_1(p;{\sf l}=0)$ exhibit {\em the same  combination} of
nuclear matrix elements  as  the numerator and
the denominator in \rf{1.1}. That is:
 \br
 {\cal I}_0(p;0)&=&|{\sf
a}|^2+|{\sf b}|^2
+3(|{\sf c}|^2+|{\sf d}|^2+|{\sf e}|^2+|{\sf f}|^2),\nn\\
{\cal I}_1(p;0) &=&2\Re[{\sf ae^*}-{\sf b}({\sf
c}-\sqrt{2}{\sf d})^*/\sqrt{3}+ {\sf f}(\sqrt{2}{\sf c}+{\sf d})^*].
\nn\\
\label{3.6}\er
with
\br\label{3.7}
{\sf a} &=& \roundbra{p,0001}{V} \roundket{0001}\isim \bra{ ^1\!\mathrm{S}_0}{\hat V}\ket{ ^1\!\mathrm{S}_0},
\nn\\
{\sf b} &=&i \roundbra{p,1101}V \roundket{0001}\isim\bra{ ^3\!\mathrm{P}_0}{\hat V}\ket{ ^1\!\mathrm{S}_0} ,
\nn\\
{\sf c} &=& \roundbra{p,0110}V \roundket{0110}\isim\bra{ ^3\!\mathrm{S}_1}{\hat V}\ket{ ^3\!\mathrm{S}_1},
\nn\\
{\sf d} &=&-\roundbra{p,2110}V \roundket{0110}\isim\bra{ ^3\!\mathrm{D}_1}{\hat V}\ket{ ^3\!\mathrm{S}_1},
\nn\\
{\sf e} &=& i\roundbra{p,1010}V \roundket{0110}\isim\bra{ ^1\!\mathrm{P}_1}{\hat V}\ket{ ^3\!\mathrm{S}_1},
\nn\\
{\sf f} &=&-i \roundbra{p,1111}V \roundket{0111}\isim\bra{ ^3\!\mathrm{P}_1}{\hat V}\ket{ ^3\!\mathrm{S}_1},
\er
where the short notation
\begin{widetext}
\br
\roundbra{p,lSJT}V \roundket{0JJT}\equiv
\roundbra{p,lL=0,\lambda=l, SJT}V \roundket{{\sf n}=1,{\sf l=L}=\lambda'=0, S'=J,JT},
\label{3.8}\er
\end{widetext}
 has been used for the matrix elements in  \rf{2.19}.

Relationships between the matrix elements $\mathcal{M}(plPL\lambda SJT;\Lambda \p)$, and
 the amplitudes ${\sf a}$, ${\sf b}$,  ${\sf c}$,  ${\sf d}$,  ${\sf e}$, and   ${\sf f}$ are shown
 in the Appendix~\ref{B}. It can be seen
 that, while the derivation of the Eq. \rf{3.6} for the $1s_{1/2}$ orbital
is mainly based on  the relation
\br
&&\mathcal{M}(plPL=0,\lambda=lSJT;\Lambda \p)=(-)^{T+1}
\nn\\
&&\times(P0|10)\roundbra{p,lSJ;T}V \roundket{0JJ;T},
\label{3.9}\er
the one for the  $1p_{3/2}$ orbital  is much more involved.
In fact, in the latter case one has to consider all matrix elements $\mathcal{M}(plPL=1,\lambda SJT;\Lambda \p)$
with $l+1\ge\lambda\ge |l-1|$, each one of them containing the c.m. matrix element $(P1|11)$ and  one or two transition
amplitudes
${\sf a},\cdots,{\sf f}$.

When expressed in the framework of the OMEM, the SM matrix elements read
\br
{\sf a}&=&
\frac{1}{\sqrt{2}}\left[{ C}_1^0+{ C}_0^0-3({ S}_1^0+{ S}_0^0)\right],\nn\\
{\sf b}&=&-\frac{1}{\sqrt{2}}
({ P}_\pi^{10}+{ P}_{K_1}^{10}+{ P}_\eta^{10}+{
P}_{K_0}^{10})+\sqrt{2}(\tilde{{
P}}_{K_1^*}^{10}+
\tilde{{ P}}_{K_0^*}^{10}),\nn\\
{\sf c}&=&\frac{1}{\sqrt{2}}\left[{ S}_0^0+{ C}_0^0-3({ S}_1^0+{ C}_1^0)\right],\nn\\
{\sf d}&=&2(3{ T}_1^{20}-{ T}_0^{20}),\nn\\
{\sf e}&=&-\frac{1}{\sqrt{6}}\left[3({ P}_\pi^{10}+{ P}_{K_1}^{10}+
2\tilde{{ P}}_{K_1^*}^{10})-{ P}_\eta^{10}-{ P}_{K_0}^{10}-2
\tilde{{ P}}_{K_0^*}^{10}\right],\nn\\
{\sf f}&=&-\frac{1}{\sqrt{3}}
\left[{ P}_\pi^{10}-{ P}_{K_1}^{10}+{ P}_\eta^{10}-{ P}_{K_0}^{10}
\right].
\label{3.10}\er
The  radial matrix elements $ {S,C,T,P,\tilde{{ P}}}$ are defined in
the Appendix~\ref{A}, and
are related to  those defined Ref.~\cite{Ba02}, namely,
 $ {\sf S,C,T,P,\tilde{{\sf P}}}$,
as ${\sf S}=S(P0|10)$,\etc. As indicated in the same appendix
 the subindices refer
to isospin and the superindices to angular momentum transitions.

Although
the SM leads to a
NOSK-like expression for $a_\Lambda$ within  the s-WA,
 both for $_\Lambda^{5}$He and $_\Lambda^{12}$C,
there are several  differences between the  SM  matrix elements and those in the  NOSK
formula, and  this is the reason for the symbol $\isim $ in \rf{3.7}:

1) The  first ones depend on the
relative momentum $p$, and the second ones do not.

2) $a,\cdots, f$~ in Eq.~\rf{1.1} are  in units of MeV$^{-2}$,
while ${\sf a},\cdots, {\sf f}$~ in Eq.~\rf{3.2}  are in units of~
MeV$^{-1/2}$. This is due to the fact that the radial wave functions for the initial states
are different.

3) As pointed out in Ref. \cite{Ba03}, they differ as well by the phase factor
$(-)^{S+J} \, i^{-l} \,$ that appears in \rf{3.7},
where the first correction is due to the change in ordering in the
Clebsch-Gordan couplings for the spins, and the second one, to the
fact that we do not include the phase $i^{l}$  in the final
partial-wave radial function.

There are still two, at first glance, quite important differences between
the NOSK formula \rf{1.1} and our SM result. They come
from the presence of the spherical harmonic
 and the integration in the Eq.  \rf{3.5}. Thus, to make them
 still more similar with each other, a few further approximations, which we feel are
 physically  quite sound, are done:

i) We assume that the $\e$'s do not play a
significant role in \rf{2.15}. Thus,    the liberated
energy $\Delta_{\jp}$ is  approximated by $\Delta=M_\Lambda-M_N$, which means
that in Eq.  \rf{3.5} is $ dU_{s_{1/2}}=dU_{p_{3/2}}$.

ii) The decay is basically back to back; therefore $\theta_p\cong 0$, and:
\be
Y_{1,0}(\theta_p,0)\cong  Y_{1, 0}(0,0)=\sqrt{{3}/{4\pi}}.
\label{3.11}\ee

iii)  The amplitudes
 ${\cal I}_\kappa(p;0)$ can be computed  at
$p\cong p_\Delta=\sqrt{M_N\Delta}$ ($P\cong 0$)
and  factored  out of the integrals.
We end up with
\be
\omega_\kappa=\hat{\kappa}
{\cal I}_\kappa(p=p_\Delta;0)\sum_{\sf L}\J_{\sf L}, \label{3.12}\ee
where
\be
\J_{\sf L}
=\frac
{2M_N}{\pi}\
\int_0^{P_\Delta}P^2\sqrt{P_\Delta^2-P^2}\O(P;{\sf L})dP,
\label{3.13}\ee
and $P_\Delta=2\sqrt{M_N\Delta}=815$ MeV.
The essential point here is that, as shown in Eq.~\rf{2.19}, the c.m.
overlaps $\O(P;L)$ have a Gaussian behaviour in the variable
$P$, and consequently the phase-space factors $P^2\,
\sqrt{P_\Delta^2 - P^2}\O(P;{\sf L})$ in~\rf{3.13} are rather narrow
peaks at $\sim 200$ MeV. On the other hand,  we have tested
numerically that the amplitudes
 ${\cal I}_\kappa(p;0)$ have a very smooth dependence on $P$ in
the range $0\le P \le 300$ MeV ($p_\Delta \ge p\ge 380$�~MeV).

Finally, the integrals $\sum_{\sf L}\J_{\sf L}$
cancel out in the numerator and the denominator in \rf{2.17},
and  we obtain
\begin{equation}\label{3.14}
a_\Lambda=\sqrt{3}
\frac{{\cal I}_1(p=p_\Delta;0)}
{{\cal I}_0(p=p_\Delta;0)},
\end{equation}
which is the NOSK-like formula that  we have been searching for.

We have also shown that
\be
 \J_0\cong \J_1\cong 2M_Np_\Delta,
  \label{3.15}\ee
 which is consistent with the result \cite[Eq.(5.3)]{Ba02}, and
 together with \rf{3.3}, reveals that within the s-WA:
\be
\omega_\kappa(s_{1/2})\cong\omega_\kappa(p_{3/2}).
\label{3.16}\ee
That is,  the $s_{1/2}$ and  $p_{3/2}$ states contribute roughly by the same
amounts, for both the proton-induced decay rate  $\Gamma_p\equiv \omega_0$ and the
numerator $\omega_1$ in the Eq. \rf{2.17}.
It is worth noting that this is not valid in the case of the neutron-induced decay rate $\Gamma_n$
where, due to the Pauli Principle, the $1s_{1/2}$-state contribution
is always larger \cite{Ba02}
than that of  the $1p_{3/2}$-state.

For the sake of consistence, the proton-induced decay rate  has to
be evaluated from
\br
&&\Gamma_p\equiv \omega_0=
\left\{
\begin{array}{ll}2M_Np_\Delta {\cal I}_0(p=p_\Delta;{\sf l}=0)
&\;\;\mbox{for ��~~~  $_\Lambda^{5}$He}            \;,\\\\
 4M_Np_\Delta {\cal I}_0(p=p_\Delta;{\sf l}=0)
&\;\;\mbox{for ��~~~  $_\Lambda^{12}$C}     \;,\\
\end{array}\right.
\label{3.17}\er
when the expression \rf{3.14} is used for the asymmetry parameter.
Note that the above result is a simple explanation  for why
\be
\Gamma_p(_\Lambda^{12}C)\cong 2 \Gamma_p(_\Lambda^{5}He).
 \label{3.18}\ee
We would like to stress that this is a purely kinematical result, and
therefore it  doesn't depend on the dynamics involved in the NMWD
process.


As an application  of the formalism developed here we exhibit the results for
$a_\Lambda$ within the simple one-pion exchange model (OPEM)
and within the ${\pi+K}$ model.
Employing  the Eqs. \rf{3.14}, \rf{A1} and
\rf{A2}, we obtain, respectively:
\be a_\Lambda^{\pi}=
-\frac{2[\left(2{T}^{20}_\pi-{S}^{0}_\pi\right) {P}^{10}_\pi]}
{18({T}^{20}_\pi)^2+({P}^{10}_\pi)^2
+3({S}^0_\pi)^2}\label{3.19}\ee
and
\begin{widetext}
\br
a_\Lambda^{\pi+K}&=&-\frac{2[2{
(3{T}_1^{20}-{T}_{K_0}^{20})P}_\pi^{10}-3({P}_\pi^{10}-{P}_{K_0}^{10})
{S}_1^0-(2{P}_\pi^{10}+3{P}_{K_1}^{10}){S}_{K_0}^0]}
{6(3{T}_1^{20}-{T}_{K_0}^{20})^2+3({P}_\pi^{10})^2+2{P}_\pi^{10}(2{P}_{K_1}^{10}-{P}_{K_0}^{10})
+({P}_{K_0}^{10})^2+9({S}_1^0)^2+3({P}_{K_1}^{10})^2
+3({S}_{K_0}^0)^2},
\label{3.20}\er
\end{widetext}
where  ${T}^{20}_1={T}^{20}_\pi+{T}^{20}_{K_1}$ and ${S}^{0}_1={S}^{0}_\pi+{S}^{0}_{K_1}$.
We remark that these results are valid for both $^5_\Lambda He$ and  $^{12}_\Lambda
C$.

\section {NUMERICAL RESULTS}

\begin{table}[h]
\begin{center}
\caption{Exact ($a_\Lambda^{exact}$) and approximate
($a_\Lambda^{approx}$) results for the symmetry parameter
$a_\Lambda$ in $^{5}_{\protect\phantom{1}\Lambda}$He.
$a_\Lambda^{exact}$ is evaluated from Eqs. \rf{2.17} and \rf{2.18},
and $a_\Lambda^{approx}$ from Eq. \rf{3.14}.\label{table1} }
\smallskip
\begin{tabular}{|c|c|c|}
\hline
OMEM &
$a_\Lambda^{exact}$&$a_\Lambda^{approx}$  \\
\hline
$\pi$&$-0.4354$&$-0.4351$\\
$(\pi,\eta,K$)&$-0.5652$&$-0.5852$\\
$\pi+\rho$&$-0.2449$&$-0.2665$\\
$(\pi,\eta,K)+(\rho,\omega,K^*)$&$-0.5117$&$-0.5131$\\
\hline
\end{tabular}
\end{center}
\end{table}

In  Tables \ref{table1} and \ref{table2} are compared the exact calculations for the asymmetry parameter
$a_\Lambda^{exact}$, evaluated from Eqs. \rf{2.17} and \rf{2.18}, with the approximated ones
$a_\Lambda^{approx}$,  obtained from  Eq. \rf{3.14}.
We see that the agreement between $a_\Lambda^{exact}$ and
$a_\Lambda^{approx}$
is  quite good for  both $^{5}_{\protect\phantom{1}\Lambda}$He and
$^{12}_{\protect\phantom{1}\Lambda}$C.
It can be seen  from Table \ref{table2} that the relation \rf{3.16}
is fairly well fulfilled, which in turn implies the validity of \rf{3.18}.

Next, we   briefly  discuss the results within the
OPEM and within the ${\pi+K}$ model  for
the individual matrix elements  listed  in Table \ref{table3}.
It is  seen from  this table and the Eq. \rf{3.19} that the contribution of the
scalar matrix element $S^{0}_\pi$
is small when  compared with those coming from $T^{20}_\pi$ and $P^{10}_\pi$. Thus, one gets:
\be
a_\Lambda^{\pi} \cong
-\frac{4{T}^{20}_\pi{P}^{10}_\pi}{18({T}^{20}_\pi)^2+({P}^{10}_\pi)^2},
 \label{4.1}\ee
which means that {\em $a_\Lambda$ is large and
negative  in the OPEM, due to the interplay between the PC tensor
(${T}^{20}_\pi$) and PV dipole (${P}^{10}_\pi$) matrix
elements}.

In the same way from the Table \ref{table3} one can easily see that
the Eq.  \rf{3.20} can be approximated as:
\begin{table*}
\caption{
Exact ($a_\Lambda^{exact}$) and approximate ($a_\Lambda^{approx}$)
results for the symmetry parameter $a_\Lambda$
in $^{12}_{\protect\phantom{1}\Lambda}$C $a_\Lambda^{exact}$ is
evaluated from Eqs. \rf{2.17} and \rf{2.18}, and $a_\Lambda^{approx}$ from
Eq. \rf{3.14}.
The HOC are  given in \rf{A3} and \rf{A4}.
\label{table2} }
\begin{ruledtabular}
\begin{tabular}{|c|ccccc|c|}
\hline
Approximation &
$\omega_0(1s_{1/2})$ &
$\omega_0(1p_{3/2})$ &
$\omega_1(1s_{1/2})$ &
$\omega_1(1p_{3/2})$ &
$a_\Lambda^{exact}$&$a_\Lambda^{approx}$ \\
\hline
$\pi$&&&&&&\\
with HOC&$0.4111$&$0.4724$&$-0.1830$&$-0.1990$&$-0.4324$&\\
without HOC&$0.4111$&$0.4327$&$-0.1830$&$-0.1863$&$-0.4377$&$-0.4501$\\
$(\pi,\eta,K$)&&&&&&\\
with HOC&$0.2788$&$0.3161$&$-0.1580$&$-0.1707$&$-0.5526$&\\
without HOC&$0.2788$&$0.2811$&$-0.1580$&$-0.1569$&$-0.5624$&$-0.5860$\\
$\pi+\rho$&&&&&&\\
with HOC&$0.4138$&$0.4607$&$-0.0984$&$-0.1096$&$-0.2379$&\\
without HO&$0.4138$&$0.4220$&$-0.0984$&$-0.1020$&$-0.2398$&$-0.2554$\\
$(\pi,\eta,K)+(\rho,\omega,K^*)$&&&&&&\\
with HOC&$0.4391$&$0.4803$&$-0.2300$&$-0.2378$&$-0.5088$&\\
without HO&$0.4391$&$0.4477$&$-0.2300$&$-0.2271$&$-0.5154$&$-0.5306$\\
\end{tabular}
\end{ruledtabular}
\end{table*}
\begin{widetext}
\br
a_\Lambda^{\pi+K}
&\cong &-\frac{2[2{
(3{T}_1^{20}-{T}_{K_0}^{20})P}_\pi^{10}-3({P}_\pi^{10}-{P}_{K_0}^{10})
{S}_1^0]}
{6(3{T}_1^{20}-{T}_{K_0}^{20})^2+3({P}_\pi^{10})^2
+2{P}_\pi^{10}(2{P}_{K_1}^{10}-{P}_{K_0}^{10})
+({P}_{K_0}^{10})^2+9({S}_1^0)^2}.
\label{4.2}\er
\end{widetext}

\begin{table}[h]
\caption{ Nuclear Matrix Elements in units of MeV$^{-1/2}$.}
\label{table3}
\bigskip
\begin{tabular}{|c|r|r|}
\hline Matrix element & $^5_\Lambda He$ & $^{12}_\Lambda
C$\\
\hline
\underline{$\pi$}&&\\
${T}_\pi^{20}$&$    -3.2402$&$    -3.7132$\\
$ {P}_\pi^{10}$&$   -8.0573$&$   -10.0379$\\
$ {S}_\pi^{0}$&$0.3876$&$0.4088$\\
\hline
$a_\Lambda^{\pi}$&$    -0.4351$&$    -0.4501$\\
\hline
\underline{$K$}&&\\
${T}_{K_0}^{20}$&$      0.4010$&$ 0.4313$\\
${T}_{K_1}^{20}$&$      1.3438$&$ 1.4455$\\
${P}_{K_0}^{10}$&$      4.8025$&$ 5.6107$\\
${P}_{K_1}^{10}$&$      0.7388$&$ 0.8632$\\
${S}_{K_0}^{0}$&$      0.1571$&$ 0.2238$\\
${S}_{K_1}^{0}$&$      0.5265$&$ 0.7501$\\
\hline
$a_\Lambda^{\pi+K}$&$    -0.5389$&$    -0.5500$\\
\hline
\end{tabular}
\end{table}
Thus the inclusion of the kaon modifies the above picture to a great
extent. The
matrix element ${T}_\pi^{20}$ goes now into the  significantly smaller term
$({T}_1^{20}-{T}_{K_0}^{20}/3)$, which would increase $a_\Lambda$ by the factor
${T}_\pi^{20}/({T}_1^{20}-{T}_{K_0}^{20}/3)\cong 1.5$.
However, as can be seen from  Table \ref{table3},  this effect is
counterbalanced to
a great extent
by the large term  $3({P}_\pi^{10})^2$ in the denominator, which now
becomes more relevant in comparison with the tensor contribution.
The kaon dipole and scalar
contributions are also appreciable and we end up with a $a_\Lambda^{\pi+K}$ which is $\cong 25\%$ larger
than $a_\Lambda^{\pi}$. We note that, while
 the contribution of ${S}_\pi^{0}$ was neglected in $a_\Lambda^{\pi}$,
that of ${S}_1^{0}$ is retained in $a_\Lambda^{\pi+K}$ because of the
coherent contribution between ${S}_\pi^{0}$ and ${S}_{K_1}^{0}$.

By employing  Eqs. \rf{A1} and \rf{A2} similar discussions can be performed for the exchanges of other
mesons.
In particular, one sees from Tables
\ref{table1} and \ref{table2} that only the $\rho$ meson can
 diminish the value of the intrinsic $\Lambda$ asymmetry.
From the last table  it is not difficult to figure out that this comes from the destructive   interference
between the $\pi$ and $\rho$ mesons in the numerator of \rf{2.17}.

\section { SUMMARIZING CONCLUSIONS AND FINAL REMARKS}

In summary, by employing the s-WA and making use of a few plausible
assumptions,  we have succeeded in shaping
 the SM formalism for the asymmetry
parameter $a_\Lambda$
into  NOSK-like formulae \rf{3.5} and/or \rf{3.14},
 which, in contrast to Eq.~\rf{1.1},
are valid for finite hypernuclei.
 The new formalism:
i) makes the theoretical evaluation of  $a_\Lambda$
more transparent, ii) explains clearly why the one-meson exchange model
is unable to account for the  experimental data of
$^5_\Lambda$He,  and iii)  helps to advance knowledge of the NMWD in
general.

It is still an open problem whether the result
\rf{2.18}, and therefore  the  formulae
\rf{3.5}, \rf{3.14}, \rf{4.1} and \rf{4.2},
are  of general validity.
Their derivation is based on the properties of the
 single-proton spectroscopic amplitude between the
core state $\ket{J_C}$ and the final states
$\ket{J_F}$, which in
the extreme SM, adopted here, have the
 {\it  simple}  expression ~\cite[Eq.(31)]{Ba05}:
\[
\Bra{J_C}a_{\jp}^\dag\Ket{J_F}=(-)^{J_F+J_C+j_p}\hat{J}_F \,.
\]
This result  allows us to perform the analytic
summation on   $J_F$, and $a_\Lambda$
becomes independent  of the nuclear structure of the final states
(\cf~Eq.\rf{2.17}).
However, because of the Pauli Principle,
it is only valid for hypernuclei with all single-particle
proton sub-shells totally full, such as happens in
$^5_\Lambda$He   and
$^{12}_{\phantom{1}\Lambda}$C.
That  is,
we still do not know whether the Eq.  \rf{2.17}, and all the developments
presented here,
 can be used for other polarized hypernuclei,
such as $^{11}_{\phantom{1}\Lambda}$B.
Very likely it does, but this has to be proved!

Quite recently, and after the present work had been basically
finished, the Barcelona group~\cite{Ch07} has stated
that a chirally motivated $2\pi$-exchange
mechanism of  D.~Jido, E.~Oset and J.~E.~Palomar~\cite{Ji01}
 strongly affects the OMEM amplitudes ${\sf a}$ and ${\sf c}$
in the Eq. \rf{3.10}, producing in this way  results that are
consistent  within the
experimental data. Within the OMEM these two amplitudes are
negative and of similar magnitudes due to the dominance
of the central spin-isospin flipping matrix element $S_1^0$ in both of
them. We feel that for a
more thorough discussion of the interplay between the two
transition mechanisms, it might be
convenient  to  extend the  formalism developed here by incorporating
the $2\pi$-exchanges into
the Eqs. \rf{3.10} and \rf{3.20}.

Last but not least,
 the very simple form of Eqs. \rf{3.5} and \rf{3.14} suggests
that it might be possible to derive these expressions
by more elementary considerations,
 instead of performing a very heavy
Racah algebra, which has been done here.
 This would be highly desirable, but so far we have not been able
to find such a simple argument.

{\bf Acknowledgements}

Two of us (C.B. and F.K.) are members of CONICET (Argentina). One of
us (F.K.) acknowledges  the support of  FAPESP (S\~ao Paulo,  Brazil).
We are grateful to Gordana  Tadi\'c and Mahir S. Hussein for critical
reading of the manuscript.

\begin{appendix}

\section{ Formulae for ${\cal I}_\k(p;{\sf  l})$ Within the OMEM}\label{A}
The transition amplitudes \rf{3.3} and \rf{3.4} that appear in \rf{3.1} are:
\begin{widetext}
\br
{\cal I}_0(p;{\sf l}=0)&=&2
\left[6(3{  T}_1^{20}-{  T}_0^{20})^2+
3({  S}_0^0)^2+9({  S}_1^0)^2+({  C}_0^0)^2
+7({  C}_1^0)^2
-4{  C}_0^0{  C}_1^0+12{  C}_1^0{  S}_1^0\right.
\nn\\
&-&6{  C}_0^0{  S}_1^0-6{  C}_1^0{  S}_0^0
+3({  P}_\pi^{10})^2+({  P}_\eta^{10})^2
+3({  P}_{K_1}^{10})^2+({  P}_{K_0}^{10})^2
+10(\tilde{{  P}}_{K_1^*}^{10})^2
\nn\\
&+&2(\tilde{{  P}}_{K_0^*}^{10})^2
-2{  P}_\eta^{10}{  P}_{K_1}^{10}+2{  P}_\pi^{10}
(2{  P}_{K_1}^{10}-{  P}_{K_0}^{10}+4\tilde{{  P}}_{K_1^*}^{10}
-2\tilde{{  P}}_{K_0^*}^{10})
\nn\\
&+&4{  P}_{K_1}^{10}(2\tilde{{  P}}_{K_1^*}^{10}-\tilde{{
    P}}_{K_0^*}^{10})
-\left.4\tilde{{  P}}_{K_1^*}^{10}({  P}_\eta^{10}+{  P}_{K_0}^{10}
+\tilde{{  P}}_{K_0^*}^{10})
\right],
\label{A1}\er\br
{\cal I}_1(p;{\sf  l}=0)&=&\frac{4}{\sqrt{3}}
\left[
3({  P}_\pi^{10}-{  P}_{K_0}^{10}+2\tilde{{  P}}_{K_1^*}^{10})
{  S}_1^0
+(2{  P}_\pi^{10}-{  P}_\eta^{10}+3{  P}_{K_1}^{10}
+4\tilde{{  P}}_{K_1^*}^{10}-2\tilde{{  P}}_{K_0^*}^{10})
{  S}_0^0\right.\nn\\
&+&({  P}_\eta^{10}-3{  P}_{K_1}^{10}-2{  P}_{K_0}^{10}
+2\tilde{{  P}}_{K_0^*}^{10})
{  C}_1^0
-({  P}_\pi^{10}-{  P}_{K_0}^{10}
+2\tilde{{  P}}_{K_1^*}^{10}){  C}_0^0\nn\\
&-&\left.2({  P}_\pi^{10}+{  P}_\eta^{10}-\tilde{{  P}}_{K_1^*}^{10}
-\tilde{{  P}}_{K_0^*}^{10})
\left(3{  T}_1^{20}-{  T}_0^{20}\right)
\right],
\label{A2}\er
\br
{\cal I}_0(p;{\sf  l}=1)&=&
6({  S}_0^1)^2+42({  S}_1^1)^2-24{  S}_0^1{  S}_1^1
+2({  C}_0^1)^2+6({  C}_1^1)^2\nn\\
&-&24{  C}_1^1{  S}_1^1
+12{  C}_1^1{  S}_0^1+12{  C}_0^1{  S}_1^1
+\frac{6}{5}\left({  T}_0^{11}+{  T}_1^{11}\right)^2
+\frac{54}{5}\left({  T}_0^{31}+{  T}_1^{31}\right)^2\nn\\
&+&14({  P}_\pi^{21})^2+2({  P}_\eta^{21})^2
+8({  P}_{K_1}^{21})^2+\frac{4}{3}({  P}_{K_0}^{21})^2
+14(\tilde{{  P}}_{K_1^*}^{21})^2+
\frac{10}{3}(\tilde{{  P}}_{K_0^*}^{21})^2
+4{  P}_\eta^{21}{  P}_{K_1}^{21}\nn\\
&-&4{  P}_\pi^{21}(
2{  P}_\eta^{21}+2{  P}_{K_1}^{21}-{  P}_{K_0}^{21}
+4\tilde{{  P}}_{K_1^*}^{21}-2\tilde{{  P}}_{K_0^*}^{21})
+4{  P}_{K_1}^{21}(
-{  P}_{K_0}^{21}-\tilde{{  P}}_{K_1^*}^{21}
+\tilde{{  P}}_{K_0^*}^{21})\nn\\
&+&4\tilde{{  P}}_{K_1^*}^{21}(2{  P}_\eta^{21}+
{  P}_{K_0}^{21}-\tilde{{  P}}_{K_0^*}^{21})
+\frac{4}{3}{  P}_{K_0}^{21}\tilde{{  P}}_{K_0^*}^{21}
+\frac{2}{3}({  P}_{K_0}^{01})^2+6({  P}_{K_1}^{01})^2
+\frac{2}{3}(\tilde{{  P}}_{K_0^*}^{01})^2
+6(\tilde{{  P}}_{K_1^*}^{01})^2\nn\\
&-&\frac{4}{3}{  P}_{K_1}^{01}{  P}_{K_0}^{01}
-\frac{4}{3}\tilde{{  P}}_{K_0^*}^{01}({  P}_{K_0}^{01}-3
{  P}_{K_1}^{01}+3\tilde{{  P}}_{K_1^*}^{01})
+4\tilde{{  P}}_{K_1^*}^{01}({  P}_{K_0}^{01}-3{  P}_{K_1}^{01}),
\label{A3}\er
and\br
{\cal I}_1(p;{\sf  l}=1)&=&\frac{4}{3\sqrt{3}}\left\{
\right.
(9{  P}_\pi^{21}+6{  P}_\eta^{21}
+3{  P}_{K_1}^{21}+8{  P}_{K_0}^{21}
+15\tilde{{  P}}_{K_1^*}^{21}+13\tilde{{  P}}_{K_0^*}^{21})
{  S}_1^1
\nn\\
&-&(3{  P}_\eta^{21}+6{  P}_{K_1}^{21}+{  P}_{K_0}^{21}
+3\tilde{{  P}}_{K_1^*}^{21}+5\tilde{{  P}}_{K_0^*}^{21})
{  S}_0^1\nn\\
&+&\frac{1}{2}(9{  P}_\pi^{21}-3{  P}_\eta^{21}
-15{  P}_{K_1}^{21}+5{  P}_{K_0}^{21}
+6\tilde{{  P}}_{K_1^*}^{21}-2\tilde{{  P}}_{K_0^*}^{21})
\left(\frac{1}{2}({  C}_1^1+{  C}_0^1)
+\frac{2}{5}({  T}_1^{11}+{  T}_0^{11})   \right)\nn\\
&-&\frac{27}{5}(3{  P}_\pi^{21}-{  P}_\eta^{21}
-3\tilde{{  P}}_{K_1^*}^{21}+\tilde{{  P}}_{K_0^*}^{21})
({  T}_1^{31}+{  T}_0^{31})\nn\\
&+&\left.(3{  P}_{K_1}^{01}-{  P}_{K_0}^{01}
-3\tilde{{  P}}_{K_1^*}^{01}+\tilde{{  P}}_{K_0^*}^{01})
({  S}_1^1+{  S}_0^1+{  C}_1^1+{  C}_0^1
-{  T}_1^{11}-{  T}_0^{11})\right\}.
\label{A4}\er
\end{widetext}

It should be noted that the formulas  for ${\cal I}_0(p;{\sf l})$ have
been presented before in \cite[Eq.(4.19)]{Ba02}), and only the results
for ${\cal I}_1(p;{\sf l})$ are  new.
The radial matrix elements $ {S,C,T,P,\tilde{{ P}}}$
have the same meaning as the  factors ,
 $ {\sf S,C,T,P,\tilde{{\sf P}}}$ in  Ref.~\cite{Ba02}, and are
related to them as: ${\sf S}=S(P0|10)$,\etc.
Nevertheless, in order to facilitate  the reading of the paper we repeat their
definitions also in the present work.

The parity conserving nuclear matrix elements are:
\br
{  C}_M^{{  l}}(p)&=&\B'_M(p{\sf l}|f_M|1{\sf l}),\mbox{for}\hspace{.5 cm}M\,=\pi,\eta,K,\rho,\omega,K^*
\nn\\
{  S}_M^{{\sf l}}(p)&=&\B_M{(p{\sf l}|f^S_M|1{\sf l})}\x\left\{
\begin{array}{rcc}
  1&\mbox{for}&M=\pi,\eta,K\\
  2&\mbox{for}&M=\rho,\omega,K^*
\end{array}
\right.,\nn\\
{  T}_M^{l{\sf l}}(p)&=&\B_M{(pl|f^T_M|1{\sf l})}\x\left\{
\begin{array}{rcc}
  1&\mbox{for}&M=\pi,\eta,K\\
  -1&\mbox{for}&M=\rho,\omega,K^*
\end{array}
\right..
\nn\er
and the parity violating ones are:
\brn
{  P}_M^{l{\sf l}}(p)&=&\A_M{(pl|f^{(-)}_M|1{\sf l})},
\nn\\
{  Q}_M^{l{\sf l}}(p)&=&\A'_M{(pl|f^{(+)}_M|1{\sf l})}.
\label{A6}\ern
The radial form factors $(p{\sf l}|f_M|1{\sf l}), {(p{\sf l}|f^S_M|1{\sf l})}, {(pl|f^T_M|1{\sf l})}$, and ${(pl|f^{(\pm)}_M|1{\sf l})}
$, and the coupling constants $\A_M, \A_M',\B_M$, and $\B_M'$ are given in \cite{Ba02}.

The compact notations are also used:
\brn
\begin{array}{ccc}
 \underline{ \tau=0 }&& \underline{\tau=1} \\
 {  C}_0= {  C}_\omega+{  C}_{K_0}
  &;& {  C}_1= {  C}_\rho+{  C}_{K^*_1}, \\
{  S}_0= {  S}_\eta+{  S}_\omega +{  S}_{K_0}  +{
S}_{K^*_0}
  &;& {  S}_1= {  S}_\pi+{  S}_\rho+{  S}_{K_1}  +{  S}_{K^*_1}, \\
{  T}_0= {  T}_\eta+{  T}_\omega+{  T}_{K_0}  +{
T}_{K^*_0}
  &;& {  T}_1= {  T}_\pi+{  T}_\rho+{  T}_{K_1}  +{  T}_{K^*_1}, \\
\end{array}
\label{A7}\ern
for the isoscalar ($\tau=0$) and the  isovector ($\tau=1$) matrix elements, and
\brn
\widetilde{{  P}}_{\eta}&=&{  P}_{\eta}-{  Q}_{K_0^*},~~
\widetilde{{  P}}_{K_0}={  P}_{K_0}-{  Q}_{\omega},~~
\widetilde{{  P}}_{K_0^*}={  P}_{K_0^*}+{  P}_{\omega},
\nn\\
\widetilde{{  P}}_\pi&=&{  P}_\pi-{  Q}_{K_1^*},~~
\widetilde{{  P}}_{K_1}={  P}_{K_1}-{  Q}_{\rho},~~
\widetilde{{  P}}_{K_1^*}={  P}_{K_1^*}+{  P}_{\rho}.
\nn\ern

\section{Relationship between the matrix elements $\mathcal{M}(plPL\lambda SJT;\Lambda \p)$, and
 the amplitudes ${\sf a}$, ${\sf b}$,  ${\sf c}$,  ${\sf d}$,  ${\sf e}$, and   ${\sf f}$             }\label{B}

\subsection{ For ${\cal I}_\k(p;\jp=1s_{1/2},0)$:}
\br
{\cal M}(p0,P0,0001;\Lambda \p)&=&{\sf  a}(p)(P0|10),\nn\\
i{\cal M}(p1,P0,1101;\Lambda \p)&=&{\sf  b}(p)(P0|10),\nn\\
{\cal M}(p0,P0,0110;\Lambda \p)&=&-{\sf  c}(p)(P0|10),\nn\\
{\cal M}(p2,P0,2110;\Lambda \p)&=&{\sf  d}(p)(P0|10),\nn\\
i{\cal M}(p1,P0,1010;\Lambda \p)&=&-{\sf  e}(p)(P0|10),\nn\\
i{\cal M}(p1,P0,1111;\Lambda \p)&=&-{\sf  f}(p)(P0|10).
\label{B1}\er
\subsection{ For ${\cal I}_\k(p;\jp=1p_{3/2},0)$:}
\br
{\cal M}(p0,P1,1011;\Lambda \p)&=&\frac{1}{\sqrt{3}}{\sf  a}(p)(P1|11),\nn\\
{\cal M}(p0,P1,1110;\Lambda \p)&=&-\frac{1}{\sqrt{6}}{\sf  c}(p)(P1|11),\nn\\
{\cal M}(p0,P1,1120;\Lambda \p)&=&-\frac{1}{\sqrt{2}}{\sf  c}(p)(P1|11),\nn\\
i{\cal M}(p1,P1,0111;\Lambda \p)&=&\frac{1}{3}\left(\frac{{\sf  b}(p)}{\sqrt{3}}-\frac{{\sf  f}(p)}{\sqrt{2}}\right)(P1|11),\nn\\
i{\cal M}(p1,P1,1111;\Lambda \p)&=&\left(\frac{{\sf  b}(p)}{3}-\frac{{\sf  f}(p)}{2\sqrt{6}}\right)(P1|11),\nn\\
i{\cal M}(p1,P1,2111;\Lambda \p)&=&\frac{\sqrt{5}}{3}\left(\frac{{\sf  b}(p)}{\sqrt{3}}+\frac{{\sf  f}(p)}{2\sqrt{2}}\right)(P1|11),\nn\\
i{\cal M}(p1,P1,1010;\Lambda \p)&=&-\frac{1}{\sqrt{6}}{\sf  e}(p)(P1|11),\nn\\
i{\cal M}(p1,P1,2020;\Lambda \p)&=&-\frac{1}{\sqrt{2}}{\sf  e}(p)(P1|11),\nn\\
i{\cal M}(p1,P1,1121;\Lambda \p)&=&-\frac{1}{2\sqrt{2}}{\sf  f}(p)(P1|11),\nn\\
i{\cal M}(p1,P1,2121;\Lambda \p)&=&-\frac{\sqrt{3}}{2\sqrt{2}}{\sf  f}(p)(P1|11),\nn\\
 {\cal M}(p2,P1,1110;\Lambda \p)&=&\frac{1}{2\sqrt{2}}{\sf  d}(p)(P1|11),\nn\\
{\cal M}(p2,P1,2110;\Lambda \p)&=&\frac{1}{2\sqrt{2}}{\sf  d}(p)(P1|11),\nn\\
{\cal M}(p2,P1,1120;\Lambda \p)&=& \frac{1}{10\sqrt{2}}{\sf  d}(p)(P1|11),\nn\\
{\cal M}(p2,P1,2120;\Lambda \p)&=&\frac{\sqrt{21}}{5\sqrt{2}}{\sf  d}(p)(P1|11),\nn\\
{\cal M}(p2,P1,3120;\Lambda \p)&=&\frac{\sqrt{21}}{5\sqrt{2}}{\sf  d}(p)(P1|11).
\label{B2}\er

\end{appendix}

\end{document}